\documentstyle[11pt,multicol]{scrartcl}
\input epsf
\def\paper#1#2#3#4#5{#1, #2 {\bf #3} (#5) \rm #4}

\def\hlfplane{\frac 12\infty}
\def\simop{\mathop{\sim}}
\def\be{\begin{equation}}
\def\ee{\end{equation}}
\def\simop{\mathop{\sim}}

\begin{document}
\title{
  \begin{center}
    {\huge {\bf Surface properties at the Kosterlitz-Thouless transition}}
  \end{center}
}
\author{Bertrand Berche\\[0cm]
       {\normalsize Laboratoire de Physique des Mat\'eriaux,}\\[-1mm]
       {\normalsize Universit\'e Henri Poincar\'e,  Nancy 1,}\\[-1mm]
       {\normalsize F-54506 Vand\oe uvre les Nancy Cedex, France}\\[-0.3cm]}
\maketitle
\vspace{-1cm}
\begin{abstract}
  {\small
  Monte Carlo simulations of the two-dimensional $XY$ model are performed in
  a square geometry with free and mixed fixed-free boundary conditions. Using a 
  Schwarz-Christoffel conformal mapping, we deduce the
  exponent $\eta$ of the order parameter correlation
  function and its surface equivalent $\eta_\parallel$  at  the Kosterlitz-Thouless transition 
  temperature. The well known value $\eta(T_{\rm KT})=1/4$ is
  easily recovered even with systems of relatively small sizes, since the shape effects are encoded in the 
  conformal mapping. The exponent associated to the surface correlations is similarly obtained
  $\eta_1(T_{\rm KT})\simeq 0.54$.
  }
\end{abstract}
\section{Introduction}\label{I}
\begin{multicols}{2}
The two-dimensional classical $XY$ model exhibits a non standard behaviour.
Long-range order of spins with continuous symmetries is indeed forbidden according to  the 
Mermin-Wagner theorem~\cite{MerminWagner66Hohenberg67}, and thus 
there is no  spontaneous magnetisation at finite temperature.
The transition is governed by unbinding of topological defects.
The low-temperature regime was investigated  by Berezinskii~\cite{Berezinskii71} 
and the mechanism of unbinding of vortices was studied by
Kosterlitz and Thouless~\cite{KosterlitzThouless73Kosterlitz74} using 
renormalization group methods. For reviews,
see e.g. Refs.~\cite{KosterlitzThouless78Nelson83,ItzyksonDrouffe89}. 

The notations are specified below: we consider a $2d$ $XY$-model on a
square lattice with two-components spin variables 
$\vec\phi_w=(\cos\theta_w,\sin\theta_w)$ located at the sites 
$w$ of a lattice $\Lambda$ of linear extent $L$. The spins interact
through the usual nearest-neighbour ferromagnetic interaction
\be
        -\frac{H}{k_BT}=K\sum_{w}\sum_\mu\vec\phi_w\cdot\vec\phi_{w+\hat\mu},
        \label{Ham}
\ee
where $K=J/k_BT$ and $\hat\mu$ is a unit vector in the $\mu-$direction.

At low temperature, the system is partially ordered, apart from
the existence of vortices, which appear in pairs in increasing
number with increasing temperature. In the low-temperature limit, the effect of vortices can 
be neglected and the behaviour is governed by spin-wave excitations, obtained 
after expanding the cosine. 
This harmonic approximation is justified, provided that the spin disorientation remains small, i.e.
at sufficiently low temperature.
Within this approximation, the quadratic energy in the Boltzmann factor leads 
to a Gaussian equilibrium 
distribution and the two-point correlation function becomes~\cite{Wegner67Sarma72}
\be
\langle\vec\phi_{w_1}\cdot\vec\phi_{w_2}\rangle\simeq|w_1-w_2|^{-1/2\pi K},
\ee
hence 
\be\eta(T)=k_BT/2\pi J.\label{eta-SW}\ee 
When the temperature increases, the influence of vortices becomes more prominent, producing
a deviation from the linear spin-wave contribution in equation~(\ref{eta-SW}), but the  
order parameter correlation function still
decays algebraically with an exponent $\eta(T)$ which depends on the
temperature.  The existence of  such a scale-invariant power-law decay of the correlation function 
is the signature of a continuous line of fixed points at low temperatures. In this paper we are 
interested in the end point of this critical line where the topological transition takes place:
at the transition temperature $T_{\rm KT}$ (usually called Berezinskii-Kosterlitz-Thouless critical
temperature), the pairs are broken and the system becomes completely disordered. 
This very peculiar topological transition is characterised by essential singularities 
when approaching the critical point from the high temperature phase, and at $T_{\rm KT}$
the correlation function exponent takes the value 
$\eta(T_{\rm KT})=1/4$~\cite{KosterlitzThouless73Kosterlitz74}.
In order to recover this result, many numerical studies were performed in the last decade using Monte 
Carlo simulations (see e.g. Refs.~\cite{FernandezEtal86GuptaEtAl88BifferalePetronzio89,Wolff89,GuptaBaillie92,JankeNather93Olsson95,Janke97,KennaIrving97}), but the analysis was made difficult by the existence of 
logarithmic corrections, e.g.
\be
  \langle\vec\phi_{w_1}\cdot\vec\phi_{w_2}\rangle
  \simop_{  L\to\infty}|w_{12}|^{-\eta}(\ln |w_{12}|)^{\eta/2},
  \label{eq9}
\ee
at $T_{\rm KT}$~\cite{ItzyksonDrouffe89}. 
Due to these logarithmic corrections which make the fits quite difficult, the value of 
$\eta$ at $T_{\rm KT}$ was a bit controversial as shown in table~1
of reference~\cite{KennaIrving97}.
The resort to large-scale simulations, up to systems as large as $L=1200$,
was then needed in order to confirm this picture~\cite{Janke97}.

Recently, we proposed a rather different approach~\cite{BercheFarinasParedes02} which is
easily implemented. We use the covariance law of  correlation
functions under the mapping of a two-dimensional system confined inside a square onto the half-infinite
plane. The scaling dimensions are then obtained through a simple power-law fit 
when the correlators are expressed as functions of  conveniently rescaled variables.
The shape effects are explicitly encoded in the conformal mapping. This is the crucial point, 
since then even very small systems lead to promising results.
We may then determine accurately the bulk correlation function exponent $\eta$ at $T_{\rm KT}$ but 
also, by choosing convenient boundary conditions, the surface critical exponent which describes 
the decay of the correlation function parallel to a free surface,
$\eta_\parallel$.

\end{multicols}\vspace{-3mm}
\section{Critical system confined inside a square with open boundaries}
\begin{multicols}{2}
For a scale-invariant  system at its critical point (which also exhibits the properties of isotropic scaling and 
short-range interactions), conformal invariance enable to include explicitly the shape
dependence in the functional expression of the correlators through the conformal
covariance transformation under a mapping $w(z)$:  
\be
\langle\vec\phi_{w_1}\cdot\vec\phi_{w_2}\rangle=|w'(z_1)|^{-x_\sigma}|w'(z_2)|^{-x_\sigma}
\langle\vec\phi_{z_1}\cdot\vec\phi_{z_2}\rangle.
\label{covconf}
\ee
The functional expression of the correlation function
inside the square geometry $w$ simply follows the mapping $w(z)$ which
realizes the conformal transformation of the half-plane $z=x+iy$ ($0\le y<\infty$) 
inside a square $w=u+iv$ of size
$L\times L$ ($-L/2\le u\le L/2$, $0\le v\le L$)
with free boundary conditions
along the four edges. This is realized by a Schwarz-Christoffel transformation~\cite{LavrentievChabat}
\be
w(z)={L\over 2{\rm K}}{\rm F}(z,k),
\quad z={\rm sn}\left({2{\rm K}w\over L}\right).
\label{eq-SchChr}
\ee
Here, $F(z,k)$ is the elliptic integral of the 
first kind, ${\rm sn}\ \! (2{\rm K}w/ L)$ 
the Jacobian elliptic sine, ${\rm K}=K(k)$ the
complete elliptic integral of the first kind, and the modulus $k$ depends on the aspect ratio
of $\Lambda$  and is here solution
of $K(k)/K(\sqrt{1-k^2})=\frac 12$.
Using the mapping~(\ref{eq-SchChr}), one obtains the local rescaling factor in 
equation~(\ref{covconf}),
$w'(z)=\frac{L}{2{\rm K}}[(1-z^2)(1-k^2z^2)]^{-1/2}$. One simply has now to include in 
equation~(\ref{covconf}) the expected behaviour of $\langle\vec\phi_{z_1}\cdot\vec\phi_{z_2}\rangle$
in the half-infinite geometry in order to get the functional expression of the corresponding
quantity in the confined system. For example in the semi-infinite geometry $z=x+iy$ 
(the free surface being defined by the $x$ axis),  the two-point correlation function is fixed up
to an unknown scaling function:  fixing one point $z_1$ close to the free surface 
($z_1=i$) of the half-infinite plane, and leaving the second point $z_2$ explore the
rest of the lattice, the following behaviour is expected:
\be
\langle \vec\phi_{z_1}\cdot\vec\phi_{z_2}\rangle_{\hlfplane}
\sim (y_1-y_2)^{-x_\sigma}\psi(\omega),\label{eq:G}
\ee
with $\eta=2x_\sigma$. The dependence on 
$\omega=\frac{y_1y_2}{\mid z_1-z_2\mid^2}$ 
of the universal scaling function $\psi$ is constrained
by the special conformal transformation~\cite{Cardy84}, and its asymptotic 
behaviour is implied by scaling e.g.  $\psi(\omega)\sim\omega^{x_\sigma^1}$ when 
$y_2\gg 1$, with $x_\sigma^1={\scriptstyle\frac 12}\eta_\parallel$
the surface scaling dimension.
Keeping $w_1$ fixed inside the square, the two-point correlation function 
becomes
\end{multicols}\vspace{-5mm}
\begin{eqnarray}  
&&\langle\vec\phi_{w_1}\cdot\vec\phi_w\rangle_{f}
\sim[\kappa(w)]^{-x_\sigma} \psi(\omega)\nonumber\\
  &&      \kappa(w)= {\rm Im}\left[{\rm sn}\frac{2Kw}{L} \right]\times
        \left| \left(1-{\rm sn}^2\frac{2Kw}{L}
        \right)  \left( 1-k^2{\rm sn}^2\frac{2Kw}{L} \right)\right|^{-1/2},
\label{eqkappa}
\end{eqnarray} 
\begin{multicols}{2}
\noindent where $f$ specifies that the boundary conditions (BCs) are chosen free along the edges of the square.
This expression is correct up to a constant amplitude determined by
$\kappa(w_1)$ which is kept fixed, but the function $\psi(\omega)$ is still
varying with the location of the second point, $w$. This  makes this expression not so useful
in practice (see e.g. reference~\cite{ChatelainBerche99} for an application).
In order to cancel the role of the unknown scaling function,
it is more convenient to work with a density profile $m(w)$ in the presence
of symmetry breaking surface fields $\vec h_{\partial\Lambda}$ 
on some parts of the boundary $\partial\Lambda$ of the lattice $\Lambda$.
In the case of fixed-free BCs denoted by $+f$ (the spins located on the $x>0$ half axis are kept fixed while 
those of the $x<0$ are free), Burkhardt and Xue have shown in the case of Ising and Potts
models that in the half-plane the order parameter profile obeys the following 
expression~\cite{BurkhardtXue91aBurkhardtXue91b}:
\be
m_{+f}(z)=y^{-\eta/2}(\cos{\scriptstyle\frac {1}{2}}\theta)^{\eta_\parallel/2}.
\ee
In the square geometry, the fixed-free BCs correspond to keeping the spins fixed for $u>0$, and
the corresponding profile is expected to obey the following ansatz:
\end{multicols}\vspace{-5mm}
\begin{eqnarray}  
&&m_{+f}(w)\sim \left(\frac{L}{2{\rm K}}\right)^{-x_\sigma}
[\kappa(w)]^{-x_\sigma}[\mu(w)]^{x_\sigma^1}, \nonumber\\
 &&\mu(w)= \frac {1}{\sqrt 2}\left(1+{\rm Re}\left[{\rm sn}\frac{2Kw}{L} \right]\times
        \left|\  \!{\rm sn}\frac{2Kw}{L}\right|^{-1}\right)^{1/2}
\label{eqkappamu}
\end{eqnarray} 
\begin{multicols}{2}
\noindent where $\kappa(w)$ was defined in equation~(\ref{eqkappa}).

In this paper we will essentially apply equation~(\ref{eqkappamu}) in order
to determine the critical exponents $\eta$ and $\eta_\parallel$ at the KT point. In 
Ref.~\cite{BercheFarinasParedes02}, the bulk exponent was deduced from a similar analysis of the order 
parameter profiles in a square-shaped system with fully fixed BCs.

\end{multicols}\vspace{-3mm}
\section{Monte Carlo simulations}
\begin{multicols}{2}
The application of the simple power-law in equation~(\ref{eqkappamu}) requires a relatively
precise numerical determination of the order parameter profile of the $2d$ $XY$ model
confined inside a square with mixed fixed-free boundary conditions. Fixing the spins along some part 
$\partial\Lambda$ of the boundary, e.g. $\vec\phi_w=(1,0), \forall w\in\partial\Lambda$,  
plays the role of ordering surface fields $\vec h_{\partial\Lambda(w)}$. 
The resort to cluster update algorithms (Wolff algorithm here) is necessary in order to prevent
the critical slowing down:
the central idea is the identification of 
clusters of sites using  bond variables connected to the spin
configuration. The spins of the clusters are then updated and
the algorithm is particularly efficient if the percolation 
threshold of the bond process occurs at the transition temperature of the spin model. 
This ensures the updating of clusters of all sizes in a single
MC sweep. For the $XY$ model, Ising variables defined in the Wolff algorithm by the sign
of the projection of the spin variables along some random direction.
The bonds are then introduced through the Kasteleyn-Fortuin random graph representation.
The percolation threshold
for these bonds coincides with the Kosterlitz-Thouless point~\cite{DukovskiMachtaChayes01}, 
which guarantees the efficiency of the
Wolff cluster updating scheme~\cite{Wolff89} at $T_{\rm KT}$. 
When one uses particular boundary
conditions, with fixed spins along some 
surface as required here, the Wolff algorithm should
become less efficient, since close to criticality the unique cluster will often
reach the boundary and no update is made in this case. To prevent this, we use the symmetry
properties of the Hamiltonian~(\ref{Ham}): even when the cluster reaches the fixed boundaries, 
it is updated, and the order parameter profile is then measured with respect to the
new direction  of the boundary
spins, $m(w)=\langle\vec\phi_w\cdot\vec \phi_{\partial\Lambda}\rangle_{\rm sq.}$.
The new configuration reached has the right statistical weight.

Using this procedure, we discard $10^6$ sweeps for thermalization, and the measurements are
performed on $10^6$ production sweeps. The simulations at $T_{\rm KT}$ on a typical system of size
$100\times 100$  takes a few hours on a standard PC with 733 MHz processor, to get a smooth profile
as shown in figure~1. 

        \epsfysize=4.5cm
        \centerline{\epsfbox{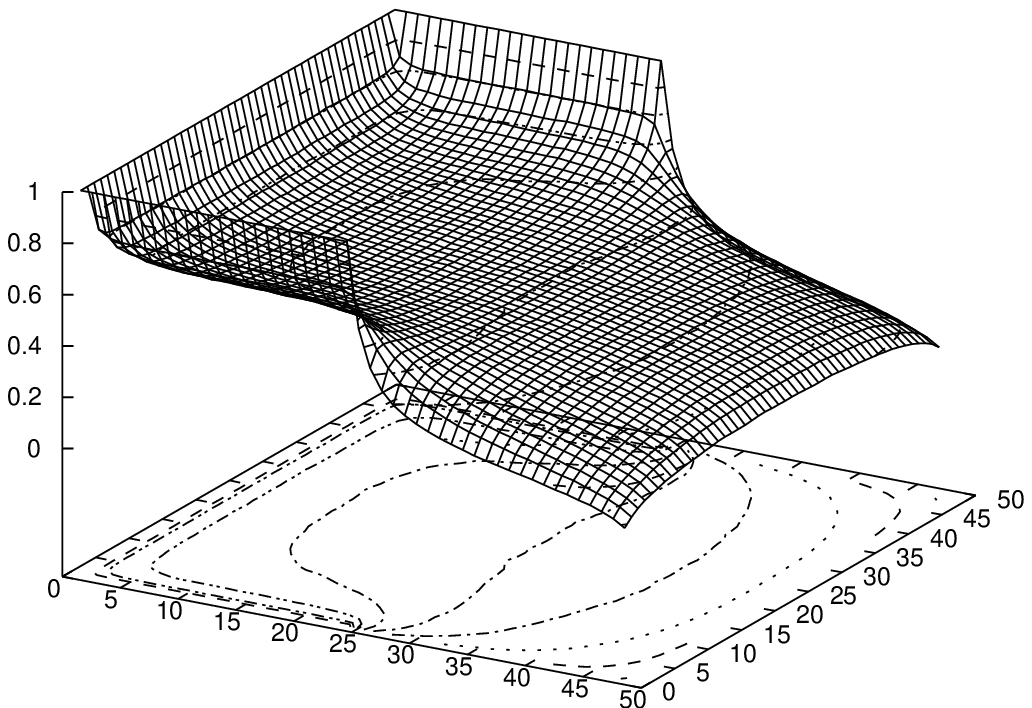}}
        {\small\noindent {\bf Figure 1:} MC simulations of the $2d$ $XY$ model inside a square
        $48\times 48$ spins  with mixed fixed-free boundary conditions at the KT
        transition temperature $k_BT_{{\rm KT}}/J=0.893$ (average over $10^6$ MCS/spin after 
        cancellation of $10^6$ for
        thermalization).}  

\end{multicols}\vspace{-3mm}
\section{Scaling of the order parameter profile $m_{+f}(w)$}
\begin{multicols}{2}
In Ref.~\cite{BercheFarinasParedes02}, we have shown that the logarithmic corrections at 
the KT point are negligible for the order parameter profile with fixed BCs. Assuming that there
will be no substantial difference for fixed-free BCs, we perform a fit of $\ln m_{+f}(w)$ against the 
two-dimensional linear expression, 
${\rm const}- {x_\sigma}\ln \kappa(w)+{x_\sigma^1}\ln\mu(w)$, as
expected from equation~(\ref{eqkappamu}). The value obtained for the bulk correlation function 
exponent $\eta$ will be a test for the absence of logarithmic corrections.
The simulations are performed at the value $k_BT_{\rm KT}/J\simeq 0.893$ found in the
literature, e.g. in Ref.~\cite{GuptaBaillie92}.
The fit leads to accurate determinations
of both exponents for different system sizes ranging between $L=20$ and $L=200$. Although it would
easily be possible to produce simulations for larger systems, there is no need here,
since the results are already quite stable as shown in table~1 and in figure~2.

\vskip4mm
{\small
{\small\noindent {\bf Table 1:} \\ Bulk and surface scaling dimensions of the order parameter at the 
Kosterlitz-Thouless transition. The values that we quote are those which correspond to a fit over all the 
lattice points (except those exactly at the boundaries) ignoring possible lattice effects.}
\vskip1mm
\begin{tabular}{rlll}
\hline\noalign{\vspace{0.4pt}}
\hline\noalign{\vspace{0.4pt}}
\hline
$L$ & $x_\sigma$ & $x_\sigma^1$ & $\chi^2/{\rm d.o.f}$ \\
\hline
20   & 0.149(33) & 0.290(15) & $0.691\times 10^{-3}$\\
24   & 0.147(34) & 0.287(15) & $0.719\times 10^{-3}$\\
28   & 0.134(22) & 0.280(12) & $1.210\times 10^{-3}$\\
36   & 0.133(35) & 0.278(12) & $1.010\times 10^{-3}$\\
48   & 0.133(35) & 0.279(10) & $0.743\times 10^{-3}$\\
64   & 0.131(35) & 0.278(10) & $0.664\times 10^{-3}$\\
100 & 0.126(17) & 0.275(14) & $0.544\times 10^{-3}$\\
200 & 0.125(23) & 0.271(11) & $0.392\times 10^{-3}$\\
\hline\noalign{\vspace{0.4pt}}
\hline\noalign{\vspace{0.4pt}}
\hline
\end{tabular}
}
\vskip4mm

This is fully coherent with the result of Kosterlitz and Thouless
RG analysis $\eta=2x_\sigma=1/4$. We can thus also consider 
as reliable the value of the surface exponent $\eta_\parallel= 2x_\sigma^1\simeq 0.54$.

\vskip8mm
        \epsfxsize=7.2cm
        \centerline{\epsfbox{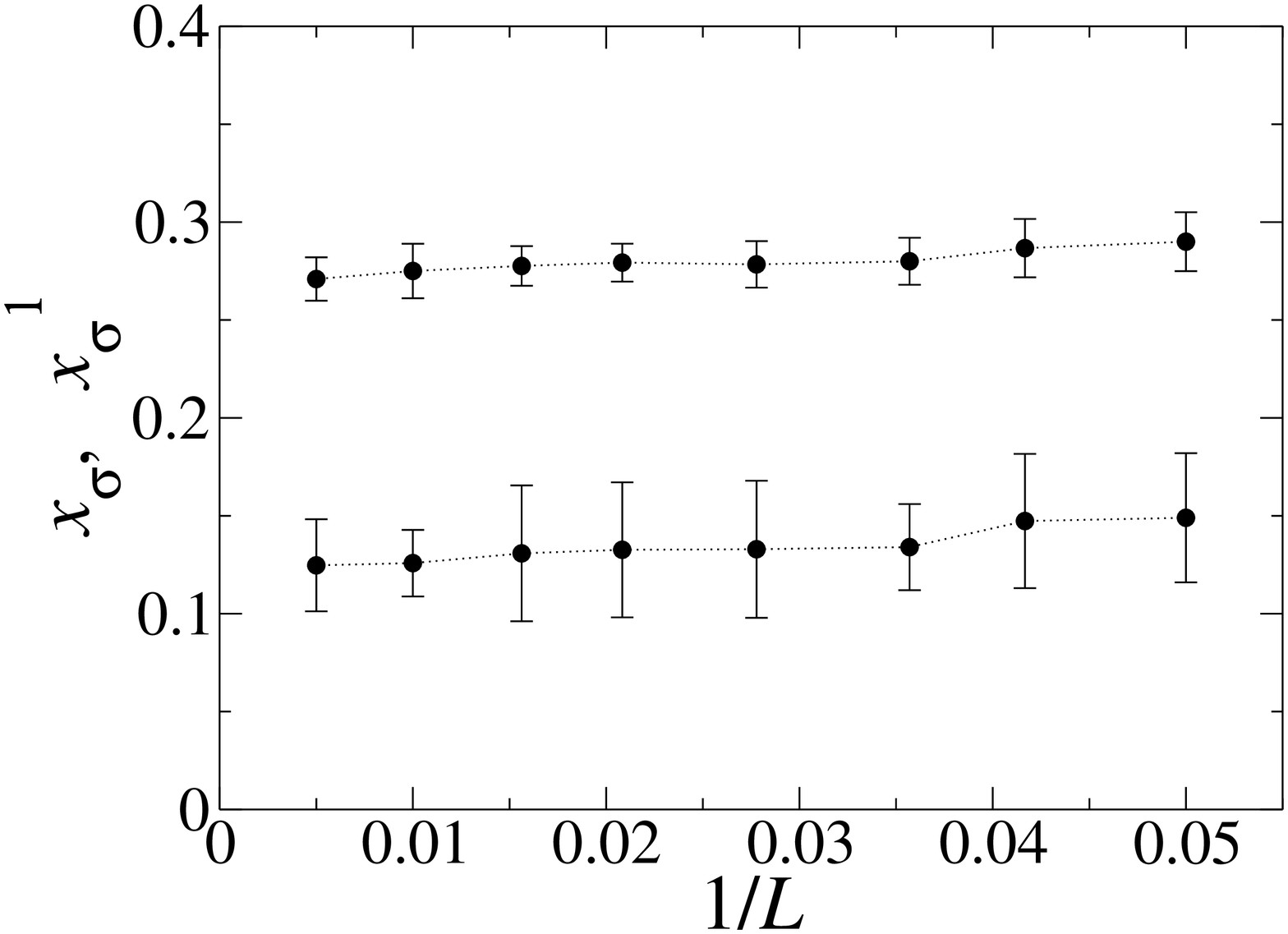}}
        {\small\noindent {\bf Figure 2:} Size dependence of the bulk and surface scaling dimensions
        at the KT transition plotted as a function of the inverse linear size of the lattice.}  
\vskip2mm

\end{multicols}\vspace{-3mm}
\section{Scaling of the correlation function $\langle\vec\phi_{w_1}\cdot\vec\phi_w\rangle_{f}$
with free boundary conditions}
\begin{multicols}{2}
Equation~(\ref{eqkappa}) provides a test in order to control the quality of the surface exponent obtained
above. Indeed, if one plots $\langle\vec\phi_{w_1}\cdot\vec\phi_w\rangle_{\rm sq.}\times
[\kappa(w)]^{x_\sigma}$ {\it vs} $\omega$, one should obtain the universal scaling function 
$\psi(\omega)$ whose asymptotic large distance behaviour should lead to a consistent
surface exponent through the expected power-law behaviour $\omega^{x_\sigma^1}$.
\vskip8mm
        \epsfxsize=7.2cm
        \centerline{\epsfbox{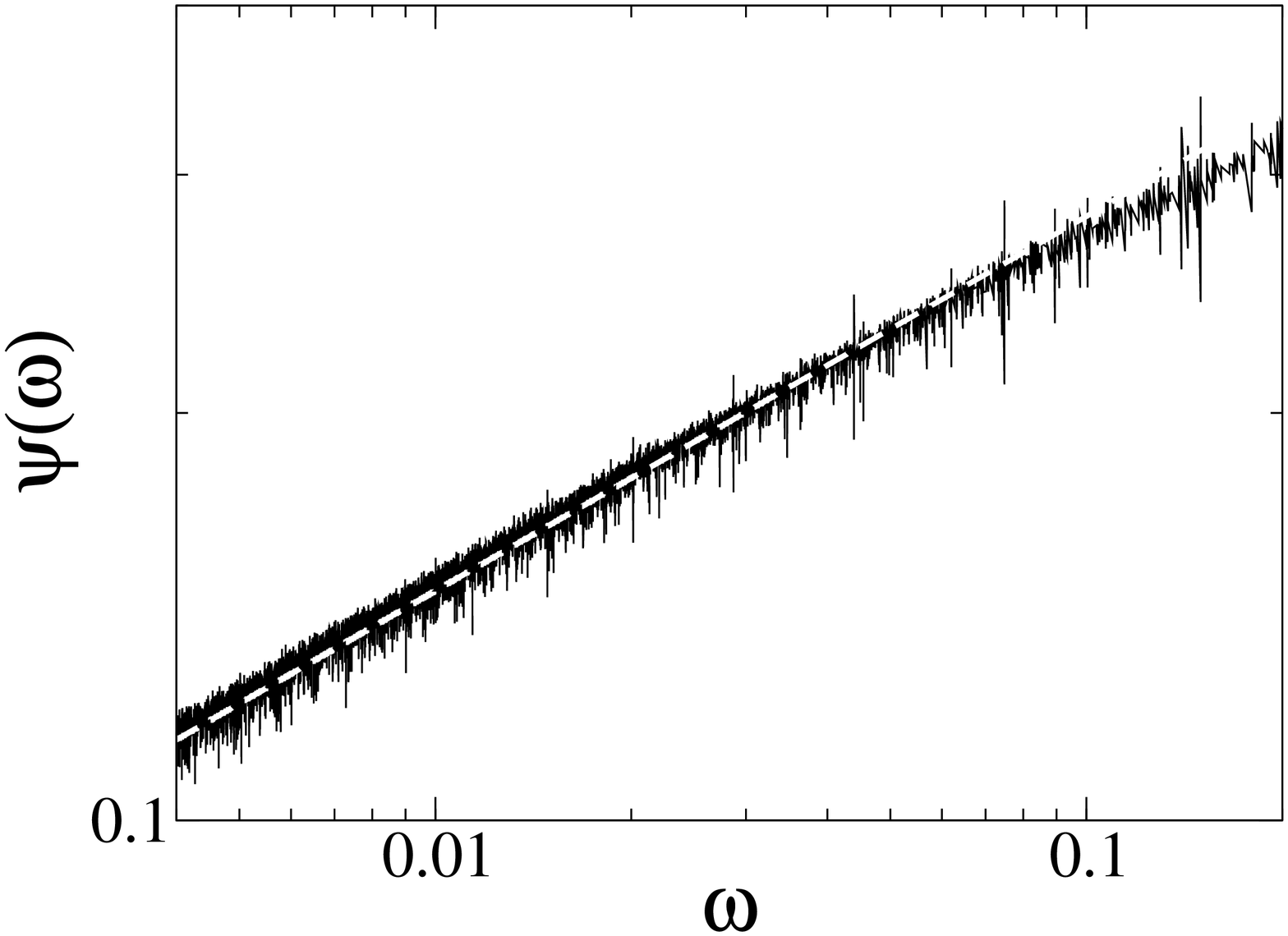}}
        {\small\noindent {\bf Figure 3:} Universal scaling function of the surface-volume correlation
        function for $L=100$. The power-law behaviour $\psi(\omega)\sim\omega^{x_\sigma^1}$ 
        shown in dotted line agrees
      with a value $x_\sigma^1\simeq 0.277$.}  
\vskip2mm
The scaling function $\psi(\omega)$ is shown in figure~3. It exhibits a clear power-law behaviour
consistent with the previous value of the surface scaling dimension. 

We mention here that the data
were obtained for both quantities, $\langle\vec\phi_{w_1}\cdot\vec\phi_w\rangle_{f}$ and $m_{+f}(w)$,
 with completely independent runs, since the boundary conditions 
are different in equations~(\ref{eqkappa}) and (\ref{eqkappamu}) and the quantities themselves
are different (2-point or 1-point correlators). 
The data collapse is quite good and supports the validity of equation~(\ref{eqkappa}).

\end{multicols}\vspace{-3mm}
\section{Conclusion}
\begin{multicols}{2}
In this paper we have performed standard MC simulations of the $2d$ $XY$ model at the 
Kosterlitz-Thouless topological transition. Combined to the use of a conformal mapping, they provide
a simple numerical determination of bulk and surface critical exponents associated to the algebraic
decay of the correlation function. The well known result of  Kosterlitz and Thouless for the bulk
exponent is recovered and as far as the author knows this is the first determination of the surface
exponent $\eta_\parallel\simeq 0.54$. 
On the contrary to the bulk case, $\eta_\parallel$ is different from
the Ising value (which is equal to $\eta_\parallel^{Ising}=1$)~\cite{Binder83}.

\end{multicols}

\begin{thebibliography}{99}
\def\And{and }
\bibitem{MerminWagner66Hohenberg67} 
  \paper{N.D. Mermin \And H. Wagner}{Phys. Rev. Lett.}{22}{1133}{1966};\ 
  \paper{P.C. Hohenberg}{Phys. Rev.}{158}{383}{1967}.
\bibitem{Berezinskii71}\paper{V.L. Berezinskii}{Sov. Phys. JETP}{32}{493}{1971}.
\bibitem{KosterlitzThouless73Kosterlitz74}
  \paper{J.M. Kosterlitz \And D.J. Thouless}{J. Phys. C}{6}{1181}{1973};\
  \paper{J.M. Kosterlitz}{J. Phys. C}{7}{1046}{1974}.
\bibitem{KosterlitzThouless78Nelson83}
  \paper{J.M. Kosterlitz \And D.J. Thouless}{Prog. Low Temp. Phys}{78}{371}{1978};\ 
  {\rm D.R. Nelson}, in {Phase Transitions and Critical Phenomena}, ed. by
  {\rm C. Domb \And J.L. Lebowitz}, Academic Press, London 1983, p.~1.
\bibitem{ItzyksonDrouffe89} {\rm C. Itzykson \And J.M. Drouffe}, {Statistical field theory},
  Cambridge University Press, Cambridge 1989, vol. 1.
\bibitem{Wegner67Sarma72} 
  \paper{F. Wegner}{Z. Phys.}{206}{465}{1967};\ 
  \paper{G. Sarma}{Solid State Comm.}{10}{1049}{1972}.
\bibitem{FernandezEtal86GuptaEtAl88BifferalePetronzio89}
  \paper{J.F. Fern\'andez, M.F. Ferreira \And J. Stankiewicz}{Phys. Rev. B}{34}{292}{1986};\ 
  \paper{R. Gupta, J. DeLapp, G.G. Batrouni, G.C. Fox, C.F. Baillie \And J. Apostolakis}
  {Phys. Rev. Lett.}{61}{1996}{1988};\ 
  \paper{L. Bifferale \And R. Petronzio}{Nucl. Phys. B}{328}{677}{1989}.
\bibitem{Wolff89}\paper{U. Wolff}{Nucl. Phys. B}{322}{759}{1989}.
\bibitem{GuptaBaillie92} \paper{R. Gupta \And C.F. Baillie}{Phys. Rev. B}{45}{2883}{1992}.
\bibitem{JankeNather93Olsson95}
  \paper{W. Janke \And K. Nather}{Phys. Rev. B}{48}{7419}{1993};\ 
  \paper{P. Olsson}{Phys. Rev. B}{52}{4526}{1995}.
\bibitem{Janke97}\paper{W. Janke}{Phys. Rev. B}{55}{3580}{1997}.
\bibitem{KennaIrving97}\paper{R. Kenna \And A.C. Irving}{Nucl. Phys. B}{485 [FS]}{583}{1997}.
\bibitem{BercheFarinasParedes02} {B. Berche, A.I. Fari$\tilde{\rm n}$as 
	Sanchez and R. Paredes}, e-print cond-mat/0208521.
\bibitem{Henkel99} {\rm M. Henkel}, {Conformal Invariance and Critical Phenomena}, 
  Springer, Heidelberg 1999.
\bibitem{LavrentievChabat} {\rm M. Lavrentiev \And B. Chabat}, {M\'ethodes
   de la th\'eorie des fonctions d'une variable complexe}, Mir, Moscou 1972, 
   Chap. VII.
\bibitem{Cardy84}  \paper{J. L. Cardy} {Nucl. Phys. B} {240 [FS12]} {514}{1984}. 
\bibitem{ChatelainBerche99} \paper{C. Chatelain \And B. Berche}{Phys. Rev. E}{60}
  {3853}{1999}.
\bibitem{BurkhardtXue91aBurkhardtXue91b}
  \paper{T.W. Burkhardt and T. Xue}{Phys. Rev. Lett.}{66}{895}{1991};\ 
  \paper{T.W. Burkhardt and T. Xue}{Nucl. Phys.}{B354}{653}{1991}.
\bibitem{refconf} Burkhard and Xue gave several predictions for conformally invariant profiles in 
  strip geometries $k+il$ or in their half-infinite counterpart $z=x+iy$:
  considering a semi-infinite system described by
        $z=x+{i}y=\rho {\rm e}^{{\rm i}\theta}$, $y\ge 0$,
        ordinary scaling implies a functional form in the half-plane, 
        $m_{ab}(z)=y^{-x_\sigma}F_{ab}(x/\rho)$, where $ab$ specifies the boundary conditions.
        Under the logarithmic transformation into a strip of width $L$,
        $\frac{L}{\pi}\ln z=k+{i}l$, one obtains 
        the order parameter profile as
        $m_{ab}(l)=\left[{L\over\pi}\sin\left(\pi{l\over L}\right)\right]^{-x_\sigma}
        F_{ab}\left( \frac{\pi l}{L}\right) $.
\bibitem{DukovskiMachtaChayes01} \paper{I. Dukovski, J. Machta \And L.V. Chayes}
        {Phys. Rev. E}{65}{026702}{2002}.
\bibitem{Binder83} K. Binder, in
        {Phase Transitions and Critical Phenomena}, edited by C.
        Domb and J.L. Lebowitz  (Academic Press, London, 1983),Vol. {8}, p. 1.
\end{thebibliography}
\end{document}